\documentclass[reprint, superscriptaddress, prl, twocolumn, amsmath, amssymb, aps, longbibliography, hidelinks]{revtex4-2}

\usepackage{graphicx}
\usepackage{soul}
\usepackage{dcolumn}
\usepackage{bm}
\usepackage{hyperref}
\usepackage{xr-hyper}
\usepackage[dvipsnames]{xcolor}
\usepackage[normalem]{ulem}

\usepackage{mathtools}
\newcommand\underrel[3][]{\mathrel{\mathop{#3}\limits_{%
			\ifx c#1\relax\mathclap{#2}\else#2\fi}}}

\newcommand{\nv}{\mathbf{n}}

\usepackage{textcomp}
\newcommand{\circled}[1]{\textcircled{\raisebox{-0.5pt}{\footnotesize #1}}}
\renewcommand{\bm}{\mathbf}
\begin{document}

\preprint{APS/123-QED}

\title{Stochastic Forces Enhance Tracer Diffusion in Non-motile Active Matter}

\author{Henry Alston}
\affiliation{Department of Mathematics, Imperial College London, South Kensington, London SW7 2AZ, United Kingdom}
\affiliation{Laboratoire de Physique, \'Ecole Normale Sup\'erieure, CNRS, PSL Universit\'e, Sorbonne Universit\'e, Universit\'e de Paris, 75005 Paris, France}


\author{Rapha\"el Voituriez}
\affiliation{Laboratoire Jean Perrin, UMR8237 CNRS, Sorbonne Universit\'e, 75005 Paris, France}
\affiliation{Laboratoire de Physique Th\'eorique de la Mati\`ere Condens\'ee, UMR7600 CNRS, Sorbonne Universit\'e, 75005 Paris, France}

\author{Thibault Bertrand}
\affiliation{Department of Mathematics, Imperial College London, South Kensington, London SW7 2AZ, United Kingdom}

\date{\today}

\begin{abstract}
\noindent Stochasticity is a defining feature of the pairwise forces governing interactions in biological systems—from molecular motors to cell-cell adhesion—yet its consequences on large-scale dynamics remain poorly understood. Here, we show that reciprocal but randomly fluctuating interactions between particles create active suspensions which can enhance the diffusion of an external tracer particle, even in the absence of self-propulsion or non-reciprocity. Starting from a lattice model with pairwise dynamics that minimally break detailed balance, we derive a coarse-grained dynamical theory for spatio-temporal density fluctuations and reveal an elevated effective temperature at short wavelengths. We then compute the self-diffusion coefficient of a tracer particle weakly coupled to our active fluid, demonstrating that purely reciprocal stochastic interactions provide a distinct and generic route to enhanced diffusivity in dense non-equilibrium suspensions.
\end{abstract}

\maketitle

Can the liquid be \textit{hotter} than the gas? Not at thermodynamic equilibrium: the diffusion of an external tracer particle generically decreases in passive suspensions of increasing density \cite{Einstein1905,Einstein1906, Einstein1911}. However, intuition from equilibrium thermodynamics routinely fails to describe the phenomena displayed in active systems \cite{Vicsek1995, Toner1995, Fily2012, Redner2013, Tjhung2018, Besse2023}. These systems are driven far-from-equilibrium by the local consumption of energy at the particle scale \cite{Marchetti2013, Bechinger2016, Tailleur2022, Bowick2022, Gompper2020} allowing microscopic dynamics which breaks detailed-balance: perhaps the clearest example is particle self-propulsion in motile active matter \cite{Seifert2005, Cates2015, Chate2020, Baconnier2025}. 

Diffusive tracers in non-equilibrium systems remain a central model for understanding transport, organization and signaling dynamics in living matter \cite{Bouchand1990, Dean2007, Bressloff2013, Bressloff2014}. The transport properties of passive tracer particles in baths of self-propelling particles have received much attention: polystyrene micron-scale beads suspended in a soap film containing motile bacteria such as  \textit{Escherichia coli} \cite{Wu2000} or microswimmers such as \textit{Chlamydomonas reinhardtii} \cite{Leptos2009} demonstrate transient ballistic (super-diffusive) motion. This was motivated by an effective {force-dipole} acting on the tracer with a finite correlation time arising due to the persistent motion of the microorganisms \cite{Wu2000} and hydrodynamic effects \cite{Leptos2009}. Moreover, passive tracers with anisotropic shape demonstrate persistent motion \cite{DiLeonardo2010, Sokolov2010} and anomalous diffusion at long times \cite{Granek2022}. 

Analytical formulations for the dynamics of a tracer can be derived through a separation of timescales argument between tracer and suspension dynamics (see Ref.\,\cite{Granek2024} for a review) or assuming weak tracer-suspension couplings \cite{Dean2011, Demery2011, Demery_2014}. The latter approach was recently employed to detail the enhancement of tracer diffusion observed in binary mixtures of particles with non-reciprocal interactions \cite{Benois2023}. While non-reciprocity provides an alternative route to making matter active \cite{AgudoCanalejo2019, Loos2020, Saha2020, Fruchart2021,Loos2022,  Alston2024_PRL, Suchanek2023}, the proposed mechanism underlying this diffusion enhancement is familiar: the formation of two-particle dimers in the suspension exhibiting transient ballistic motion \cite{Benois2023, Cocconi2023}.   

Importantly, systems can be made active without the persistent motion of individual particles (as in self-propulsion) or particle pairs (as in non-reciprocity). Indeed, stochastic reciprocal forces provide an example \cite{Alston2022b} and recent experimental works have demonstrated how they endow dense clusters with heightened response to external forces \cite{Bonazzi2018, Kuan2021, Tennenbaum2016, Oriola2022} and help direct growth during morphogenetic processes \cite{Mongera2018, Krajnc2020, Kim2021}. They have recently also been shown to drive non-trivial diffusion properties: a case in point is that of the bacterium \textit{Neisseria meningitidis} which exhibits enhanced self-diffusion in dense aggregates due to (intermittent) pili-mediated interactions \cite{Bonazzi2018}. Naturally, the underlying mechanism must deviate from the active systems discussed at the outset as the bacteria's mean-squared displacements do not exhibit transient ballistic scaling \cite{Bonazzi2018, Alston2024b}, suggesting an alternative route to activity-enhanced diffusion of passive tracers. 

\begin{figure*}
    \includegraphics[width=\linewidth]{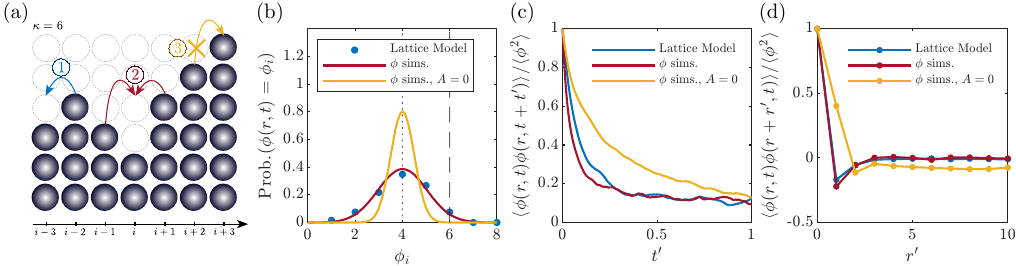}
    \caption{\textit{Lattice model and continuum description ---} (a) Schematic of the present lattice model where particles \circled{1} diffuse to neighboring sites and \circled{2} interact  over two sites by hopping to the center of mass, while \circled{3} partial exclusion ensures slowed rates on to crowded sites and enforces a finite carrying capacity, enforced here as $\kappa=6$ (dashed grey line). We then set $\bar{n}=4$ (dotted grey line) and compare the spatial and temporal correlation functions in simulations of the two dynamics (discrete lattice model and continuous field dynamics). We find quantitative agreement between (b) the distribution of particle numbers, (c) the relaxation time of the two-time correlation function and (d) the relaxation length for the two-space correlation function. Note that we set the lattice spacing to be equal to $a=1$.
    }
    \label{fig:latt_nums1}
\end{figure*}

In this Letter, we demonstrate that stochastic reciprocal forces can create active suspensions where the diffusion of an external tracer is larger in denser phases. Starting from a lattice model, we derive a linear field theory describing non-equilibrium density fluctuations due to pairwise dynamics between particles that break detailed-balance (but conserve centre-of-mass) and drive liquid-gas phase separation. These dynamics capture the essential features of particle fluctuations in dense suspensions with isotropic active stresses arising from fluctuating interactions such as pili binding-unbinding or ATP-driven actin-myosin turnover in cell monolayers. We then calculate the self-diffusion coefficient of a passive tracer weakly coupled to this fluctuating field, quantifying how the active interactions in the suspension enhance tracer diffusion.

\textit{Lattice model with minimal detailed-balance breaking attractive interactions ---} We consider a lattice model of the \textit{partial exclusion process} \cite{Tailleur2008b, Floreani2021} on a periodic domain $\mathcal{L}$ with spacing $a$, augmented with the most minimal pairwise interaction that breaks detailed-balance while enforcing center-of-mass conservation (ensuring reciprocity) and bringing particles together (attractive).

Let $n_i(t)$ denote the particle number at site $i$ and define the state $\bm{n}=\{n_i\}_{i\in \mathcal{L}}$. The model dynamics are then defined as follows:
\begin{itemize}
    \item[\circled{1}] Particles at site $i$ jump to site $i\pm 1$ with rate $\alpha_{i,\pm1}(\bm{n}) = D_r n_i (1- n_{i\pm1}/\kappa)$, where $D_r^{-1}$ sets the typical timescale associated with the diffusive motion and $\kappa$ is the carrying capacity at each site, assumed to be constant throughout the system;
    \item[\circled{2}] Two particles at sites $\{i-1, i+1\}$ simultaneously hop to site $i$ with rate $\beta_i(\bm{n}) = k_r n_{i-1} n_{i+1} \max(0,1 - n_i / (\kappa-1))$, where the partial exclusion due to the carrying capacity takes into account the $+2$ increase in particle number. 
\end{itemize} 
Note that as $n_i\leq \kappa$ the rates introduced above are always non-negative. The breaking of detailed-balance is evident as the probability that a pair of particles hop together in the next $dt$ time units is $\mathcal{O}(dt)$, whereas the reverse event (two diffusive hops apart) happens with probability $\mathcal{O}(dt^2)$. Finally, the choice of partial exclusion prevents the unphysical collapse of all particles onto a single site. A schematic of these hopping rules is provided in Fig.\,\ref{fig:latt_nums1}(a). 

Our choice of lattice hopping rules captures, at a most minimal level, pairwise dynamics due to intermittent attractive forces: particle pairs diffuse apart before hopping together stochastically in a manner that breaks detailed balance \cite{Alston2022b}. These forces were previously studied in an off-lattice, particle-based model: the deterministic density fluctuations for this system were shown to pertain to the non-equilibrium phase separation field theory \textit{Active Model B+} and to drive the coexistence of finite size clusters in the long time limit \cite{Alston2022a}. By using an on-lattice model, our current approach provides an exact microscopic derivation of the stochastic terms appearing in the field theory, going beyond the coarse-grained treatment of Ref.\,\cite{Alston2022a}. This framework already breaks fluctuation–dissipation at the linear level, leading to non-equilibrium corrections to the self-diffusion of a tracer particle to linear order (as we show formally below).

\textit{Dynamical description for the particle number fluctuations ---} We now coarse-grain the dynamics of the lattice model following the approach of Ref.\,\cite{Lefevre2007}. We remark here that while providing accurate results for a range of models, the coarse-graining procedure was shown to provide spurious phase separation results for asymmetric interaction kernels \cite{Thompson2011}. Therefore, we compare our results below to numerical simulations of the underlying process to identify when our derived dynamics suitably captures those of our lattice model.  

Following Ref.\,\cite{Lefevre2007}, we derive a Langevin equation for the particle number field through the construction of a Martin-Siggia-Rose dynamical action. The procedure involves identifying the action $\mathcal{S}(\nv(t), \tilde{\nv}(t))$ from which the path probability for the process can be evaluated as 
\begin{equation}
    \mathbb{P}(\nv(t)) = \frac{1}{Z}\int \mathcal{D}\tilde{\nv}~e^{-\mathcal{S}(\nv(t), \tilde{\nv}(t))}
\end{equation}
where $\nv(t) = \{n_i(t)\}_{i \in \mathcal{L}}$, $\tilde{\nv}= \{\tilde{n}_i(t)\}_{i \in \mathcal{L}}$, $\mathcal{D}\tilde{\nv} = \prod_{i\in\mathcal{L}} d\tilde{n}_i$ and $Z$ normalizes the distribution. The action can be determined (see full details in \footnote{See Supplementary Material at [...].}\label{Note1}) and takes the form 
\begin{align}
    \mathcal{S}(\nv, \tilde{\nv}) = \sum_{i \in \mathcal{L}} \int dt \bigg[\tilde{n}_i\partial_t n_i &- \sum_{\nu=\pm 1}  \alpha_{i,\nu} \left(e^{\tilde{n}_{i+\nu}-\tilde{n}_{i}} -1\right) \nonumber\\
    								&- \beta_{i} \left(e^{-\nabla^2 \tilde{n}_{i}} - 1\right)\bigg]~.
\end{align}

To derive an equation for $\partial_t n_i$, we then re-write all non-local terms as local terms (site $i$ in the summand) using gradient (Taylor) expansions, then expand the exponential terms involving the conjugate fields. We then follow the same approach taken for other lattice models with simultaneous particle hopping and fix the lattice spacing to $1$ \cite{Hexner2017, Han2024} and make the (\textit{a priori} uncontrolled) assumption that we can discard higher-order gradient terms on the basis that they are dominated by leading-order terms. In the current work, we only seek to capture the particle number fluctuations $\phi(x, t) \equiv n(x, t) - \bar{n}$, such that, under the assumption that $\phi\ll \bar{n}$, we only need to keep leading-order terms in $\phi$. Note that keeping higher-order terms would be necessary to derive an accurate dynamics for the particle number field $n(x, t)$ to investigate phase-coexistence densities, for instance. 

At this stage, a dynamical equation for $\phi$ describing Gaussian fluctuations about the mean-field fixed point can be identified from the action: the $\mathcal{O}(\tilde{n})$ terms contributing to the deterministic dynamics and $\mathcal{O}(\tilde{n}^2)$ terms to stochastic terms. The resulting dynamics is given by
\begin{equation}\label{eq:phieq}
    \partial_t \phi = D_0\partial_x^2 \phi - \gamma \partial_x^4 \phi + \sqrt{2D_1} \partial_x{\Lambda} + \sqrt{A} \partial_x^2 \xi,
\end{equation}
where crucially $\gamma$, $D_0$, $D_1$ and $A$ are all derived in terms of the microscopic jump rates and exclusion factors with the coefficients for the deterministic terms given by
\begin{subequations}
\begin{align}
D_0 &= D_r - k_r \left( 2 \bar{n} - \frac{3\bar{n}^2}{\kappa-1} \right) \\
\gamma &= \frac{1}{12}\left[D_r  + k_r \left( 14 \bar{n} - \frac{15\bar{n}^2}{\kappa-1} \right) \right]. 
\end{align}
\end{subequations}
We remark here that the decision to neglect higher-order terms restricts the validity of the dynamical description to long-wavelength dynamics: when calculating observables below, we impose a minimum wavelength (set by the lattice spacing to unity) to avoid divergences at higher Fourier modes $q$. The stochastic terms $\Lambda(x, t)$ and $\xi(x, t)$ are zero-mean, unit variance Gaussian white noise processes and have coefficients
\begin{align}
D_1 = D_r\bar{n}\left(1 - \frac{\bar{n}}{\kappa}\right)~~\mathrm{and} ~~A = k_r\bar{n}^2\left(1 - \frac{\bar{n}}{\kappa-1}\right).
\end{align}

While the dynamics are written here in 1D, the extension to higher dimensions is trivial due to the orthogonal nature of the lattice dynamics. It is straightforward to confirm that the lattice model exhibits phase separation: $D_0$ can be negative for some range of $\bar{n}<\kappa$ for sufficiently large $k_r$, indicating the presence of a spinodal instability (as captured in the off-lattice model of Ref.\,\cite{Alston2022a}). We consider densities $\bar{n}$ above this range such that $D_0(\bar{n})>0$ and the linearized dynamics are stable.

We compare the particle number fluctuations in the lattice model to those described by Eq.\,\eqref{eq:phieq} by simulating both and comparing the spatial and temporal correlation functions. We simulate the lattice model dynamics using the Gillespie algorithm \cite{Gillespie1977} and solve the discretized version of Eq.\,\eqref{eq:phieq} with unit lattice spacing using centered finite-difference operators for spatial gradients and an explicit Euler method for time integration. The results are given in Fig.\,\ref{fig:latt_nums1}(b-d): we find quantitative agreement in the temporal and spatial relaxation of fluctuations. Interestingly, the spatial correlations in panel (d) show a depletion at $r'=1$ site, compared to the equilibrium picture for attractive interactions (in yellow). This arises due to the stochastic nature of the interactions and represents a non-equilibrium feature of our model. We remark here that $D_1$ can be interpreted with a thermal temperature: the self-diffusion coefficient of an isolated particle in the lattice dynamics studied is $D_r$ (for unit lattice spacing), effectively setting the temperature $T$ of the system. With this in mind, we identify $D_1 = \mu_\phi T$, the product of the field motility, $\mu_\phi = \bar{n}(1-\bar{n}/\kappa)$, and the temperature, $T$.

We now analyze the form of Eq.\,\eqref{eq:phieq}, identifying the presence of a second noise term with prefactor $\sqrt{A}$. This noise term vanishes for $k_r=0$, indicating that this term captures the stochastic nature of the attractive interactions. It is crucial for the agreement between lattice and field dynamics, as demonstrated in Fig.\,\ref{fig:latt_nums1}(b-d). The noise appears under a Laplacian operator: such noise terms have been identified previously in the study of hyperuniform systems \cite{Torquato2018} whereby interactions conserving the center-of-mass drive the suppression of low wavenumber (large-scale) density fluctuations \cite{Hexner2015, Hexner2017, Galliano2023}. This manifestation of long-range order was also recently argued to arise due to dynamical active stresses in $d=2$ dimensions \cite{Keta2024}. The connection with the current work is immediate: in a linearized description of the particle number or density, stochastic isotropic active stresses—modeled as white noise—give rise to the noise term $\xi$ that appears in Eq.\,\eqref{eq:phieq}.

As such, we believe that the form of the dynamics in Eq.\,\eqref{eq:phieq} extend beyond the lattice model studied, in principal describing (to leading-order in deterministic and stochastic terms) symmetric particle number fluctuations in systems subject to thermal and active interaction-mediated (center-of-mass conserving) fluctuations. It is straightforward to confirm that the last term in Eq.\,\eqref{eq:phieq} is sufficient to drive a departure from equilibrium fluctuations: recalling $D_1=\mu_\phi T$, we write the dynamics in Fourier space as
\begin{subequations}
\begin{align}
    \partial_t\tilde{\phi}(q, t) &= \left[-D_0q^2 -\gamma q^4\right]\tilde{\phi} + \sqrt{\mu_\phi}\eta(q, t), \\ 
    \langle \eta(q, t)\eta(q',t')\rangle &= \left[2Tq^2 +A'q^4 \right]\delta(q+q')\delta(t-t').
\end{align}
\end{subequations}
The dynamics for the Fourier mode $q$ are exactly of \textit{Model B} form (in the Gaussian approximation) with the effective temperature $T_{\rm eff}(q)=T + A' q^2/2$ \cite{Hohenberg1977} where $A'=A/\mu_\phi$. The $q$-dependence here indicates that we cannot simply describe the dynamics for the whole field $\phi$ through a re-scaling of the temperature, indicating a genuine departure from equilibrium fluctuation dynamics. This contrasts with the top-down derivations of dynamical descriptions for particle fluctuations in active systems such as \textit{Active Model B} \cite{Wittkowski2014} and \textit{B+} \cite{Tjhung2018, Fausti2021, Besse2023}, where higher-order deterministic terms are introduced to drive a departure from equilibrium dynamics.

\textit{Self-diffusion coefficient of tracer particle coupled to field dynamics Eq.\,\eqref{eq:phieq} in $d$ dimensions ---} We now employ the dynamical framework we derive above to study the effect of stochastic interactions on the diffusivity of an external, Brownian tracer particle (at temperature $T$) at position $\mathbf{y}(t)$ which we reciprocally couple to the dynamical field $\phi$. Note that in doing so we implicitly assume a seperation of length-scales between the dynamics of the particles in our active suspension (captured by the lattice model) and the tracer particle. This allows us to quantify the transport properties of our tracer particle, Thus, what we calculate below captures the transport properties of our active suspension, rather than the statistics of the random motion of the suspension particles.

 As a starting point, we consider the field dynamics from before but now written in $d$ dimensions. Upon coupling to the tracer, the dynamics takes the form:
 \begin{subequations}\begin{align}\label{eq:field-tracera}
    \partial_t \phi(\mathbf{r}, t) &= D_0\nabla^2 \phi - \gamma \nabla^4 \phi + h \nabla^2 \delta(\mathbf{y})\nonumber \\
    &\quad + \sqrt{2D_1} \nabla\cdot\mathbf{\Lambda}(\mathbf{r}, t) + \sqrt{A} \nabla^2 \xi(\bm{r},t)\\
    \dot{\mathbf{y}}(t) &= -h  \nabla \phi(\mathbf{y},t) + \sqrt{2D_y}\pmb{\eta}_y(t) \label{eq:field-tracerb}
\end{align}\end{subequations}

\begin{figure}
    \centering
    \includegraphics[width=0.9\linewidth]{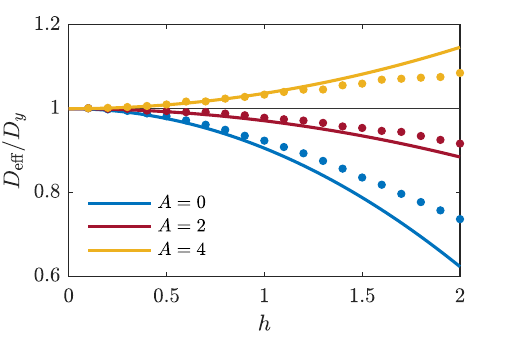}
    \caption{\textit{Range of validity for tracer self-diffusion coefficient result ---} We compare results from simulations of Eqs.\,\eqref{eq:field-tracera} and \eqref{eq:field-tracerb} (black dots) to our analytic result Eq.\,\eqref{eq:Defftracer} for the self-diffusion coefficient of the tracer particle. We find accurate agreement up to $h=1$ and confirm the observation that sufficiently strong interaction fluctuations can enhance the diffusion of the tracer. Parameter values are $D_0=\gamma=D_1=1$.}
    \label{fig:tracer_Deff}
\end{figure}

Analytical frameworks for calculating statistics of the tracer particles motion for these particle-field coupled dynamics have been established previously \cite{Dean2011, Demery2011, Demery_2014}. Here, we extend these approaches to field dynamics augmented with noise terms breaking fluctuation-dissipation through gradient terms (or equivalently an effective temperature for the field that is dependent on the Fourier mode $q$). A full re-derivation in this case is given in \cite{Note1}. The result for the self-diffusion coefficient is then calculated from the long-time mean-squared displacement:
\begin{equation}
    D_{\rm eff} = \lim_{t\rightarrow \infty} \frac{\langle |\mathbf{y}(t)-\mathbf{y}(0)|^2\rangle}{2dt}.
\end{equation}
The final result for the coupled dynamics of Eqs.\,\eqref{eq:field-tracera} and \eqref{eq:field-tracerb} to quadratic order in $h$ takes the form
\begin{align}\label{eq:Defftracer}
    \frac{D_{\rm eff}}{D_y} - 1 &= \\
    - &\frac{h^2}{d} \int \frac{d^d\mathbf{q}}{(2\pi)^d} \frac{\mu_\phi T_{\rm eff}(\mathbf{q}) + (2-D_y^{-1}\mu_\phi T_{\rm eff}(\mathbf{q}))\Delta(\mathbf{q})}{\Delta(\mathbf{q})(D_y + \Delta(\mathbf{q}))^2}\nonumber
\end{align}
where $D_0$ is taken to be positive and we have defined $\Delta(\mathbf{q}) = D_0+\gamma |\mathbf{q}|^2$ and recall the definition $\mu_\phi T_{\rm eff}(\mathbf{q})=D_1 + \frac{A}{2}|\mathbf{q}|^2$. 
We explore the range of validity of this perturbative result by comparing it to numerical simulations of the particle-field dynamics Eqs.\,\eqref{eq:field-tracera} and \eqref{eq:field-tracerb}. For comparison with the lattice model studied above, we recall that we impose a finite lattice spacing (set to unity) in deriving the continuum description, thus imposing a UV cutoff for the Fourier modes: $q_{\rm max}=2\pi$. The numerical method is detailed in \cite{Note1} and the results are given in Fig.\,\ref{fig:tracer_Deff}, where we find a good agreement for $h<1$.

From this analytic result, we draw two important conclusions. First, we conclude that $A>0$ does not generically lead to an enhancement in the self-diffusion of the tracer particle. To see this, consider the terms in the numerator of the integrand in Eq.\,\eqref{eq:Defftracer} which are proportional to $A$: we can easily see that in the case where $-1+D_0+\gamma q_{\rm max}^2 < 0$, then the correction due to $A>0$ is negative (leading to $D_{\rm eff}/D_y<1$ for all values of $h$).

More strikingly, in the case where $D_0$ and $\gamma$ are such that this correction is positive, sufficiently strong interaction fluctuations can drive an enhancement of the self-diffusion of the tracer to the point where it becomes more diffusive due to coupling in the field, that is, $D_{\rm eff}>D_y$ for sufficiently large $A>0.$ This represents a truly non-equilibrium consequence of our model: studies of field dynamics satisfying fluctuation-dissipation universally report a decrease in tracer self-diffusion upon weak (reciprocal) coupling to an external field \footnote{The case where the coupling is not reciprocal, i.e.\,the field does not feel the force of the particle, can exhibit increases in the self-diffusion of the tracer upon coupling \cite{Dean2011}\label{Note2}}. Here, we demonstrate how non-equilibrium fluctuations, driven by microscopic reciprocal interactions breaking detailed balance, can drive an enhancement in these transport properties of the tracer. We remark here that we can also consider the effective motility $\kappa_{\rm eff}$ of a tracer particle being dragged through with some force $\mathbf{f}$, defined as $ \kappa_{\rm eff}(\mathbf{f}) = \lim_{t\rightarrow\infty}\left({\langle \mathbf{y}(t)\rangle}/{\mathbf{f}t}\right).$ However, as we show in \cite{Note1}, the effective motility is unaffected by $A$ and is thus identical to that of Model B in the Gaussian approximation, as studied in Ref.\,\cite{Dean2011}.

\begin{figure}
    \centering
    \includegraphics[width=0.9\linewidth]{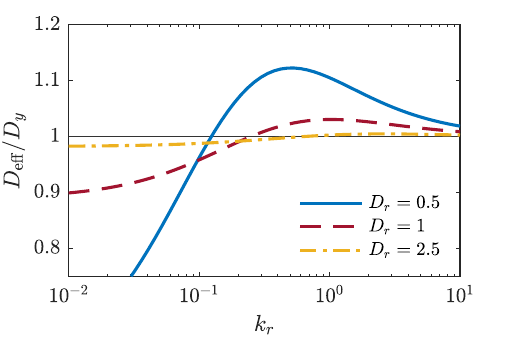}
     \caption{\textit{Connecting tracer diffusivity enhancement to microscopic hopping rates --- }We plot the analytic result Eq.\,\eqref{eq:Defftracer} where the constants are evaluated using the expressions for the particle model with $\bar{n}=4$ and $\kappa=6$ as in Fig.\,\ref{fig:latt_nums1} and $h=0.5$. We observe a range for the interaction rate $k_r$ in which the diffusion of the tracer particle is enhanced, including a finite $k_r$ that maximises this effect.
     }
    \label{fig:micro}
\end{figure}

\textit{Relation to the microscopic jump rates  ---} Finally, we relate our analytic result Eq.\,(\ref{eq:Defftracer}) to our original lattice model to intepret this diffusion enhancement in terms of microscopic timescales (those of self-diffusive and pairwise dynamics): the results are plotted in Fig.\,\ref{fig:micro}. We identify three regimes: for $D_r \gg k_r$, the detailed-balance breaking lattice dynamics are unimportant and the result mirrors that of Refs.\,\cite{Dean2011, Demery2011} for a field dynamics of \textit{Model B} form where the field and the tracer evolve on comparable timescales. For $D_r \ll k_r$, the timescales for the tracer and field dynamics are widely separated: in this limit, only the $k_r$ terms are important in the field dynamics, so $k_r$ sets the timescale for $\partial_t\phi$, whereas the dynamics for the tracer do not scale with $k_r$. In this case, $D_{\rm eff}\rightarrow D_y$ as $k_r\rightarrow\infty$ as observed previously for fast field dynamics \cite{Dean2011, Demery2011}.

In Fig.\,\ref{fig:micro}, we observe that tracer diffusion enhancement requires the interaction and diffusion rates to be comparable and is optimised for finite interaction rate, reminiscent of stochastic resonance \cite{Jung1991, Gammaitoni1998, Wellens2004}. We note here that the observed enhancement strongly resembles that of a particle coupled non-reciprocally to the field: tracer diffusive enhancement is observed when the field does not deform due to the presence of the particle (so-called ``passive" rather than ``active" tracers \cite{Note2}). This is also optimised for a finite relaxation time of the field dynamics. We propose here that the non-equilibrium fluctuations, driven by the second ($\propto\sqrt{A}$) noise term, may combat the deformation of the field due to the tracer preventing local trapping, hence why our full system may resemble one where the field does not see the particle. 

\textit{Discussion and conclusion ---} We have demonstrated that a minimal example of attractive, reciprocal and detailed-balance-breaking pairwise interactions are sufficient to drive non-equilibrium fluctuations in particle number. After a mean-field coarse-graining procedure, we confirmed that the derived dynamics for the particle number fluctuations quantitatively capture those of the underlying model when the system is dense. Moreover, the general form of our continuum description Eq.\,\eqref{eq:phieq} describes a broad class of non-motile active systems: we consider only the minimal ingredients of diffusion augmented with center-of-mass conserving stochastic interactions \cite{Hexner2015, Hexner2017} breaking detailed-balance. Our equation also arises from a hydrodynamic description of active systems with fluctuating isotropic stresses \cite{Keta2024} modelling type-IV pili mediated interactions between \textit{Neisseria meningitidis} \cite{Bonazzi2018, Kuan2021, Tennenbaum2016, Oriola2022} and dynamical cell-cell tensions driven by actin-myosin turnover in embryonic cell tissues in zebrafish \cite{Mongera2018, Krajnc2020, Kim2021}.

We then calculated the self-diffusion coefficient for a tracer particle coupled to the fluctuating field dynamics, deriving an analytic expression to quadratic order in the coupling strength $h$ (which we compared to results from numerical simulations, concluding on good agreement up to $h\approx 1$). We illustrated how coupling to the field can enhance the diffusive motion of the tracer, an effect that is inimitable for passive scalar field dynamics with reciprocal tracer-field couplings \cite{Dean2011, Demery2011}. 

We recall here that in a previous work \cite{Alston2022a}, intermittent attractive forces were studied in an off-lattice, particle-based model. Developing a framework extending such analysis to the model of Ref.\,\cite{Alston2022a} remains an open problem that we hope to address in future work. More generally, we believe the results of this work further establish stochastic pairwise forces as a genuine source of activity, whose consequences for emergent collective phenomena warrant greater attention in the field of active matter.

\textit{Acknowledgements --- }The authors are grateful to Vincent D\'emery, Julien Tailleur, Paul C. Bressloff, Sunghan Ro and Emir Sezik for interesting discussions. H.\,A.\,was supported by a Roth PhD scholarship funded by the Department of Mathematics at Imperial College London. 


%

\end{document}


\preprint{APS/123-QED}

\title{Supplementary Material: Stochastic Forces Enhance Tracer Diffusion in Non-motile Active Matter}

\author{Henry Alston}
\affiliation{Department of Mathematics, Imperial College London, South Kensington, London SW7 2AZ, United Kingdom}
\affiliation{Laboratoire de Physique, \'Ecole Normale Sup\'erieure, CNRS, PSL Universit\'e, Sorbonne Universit\'e, Universit\'e de Paris, 75005 Paris, France}

\author{Rapha\"el Voituriez}
\affiliation{Laboratoire Jean Perrin, UMR8237 CNRS, Sorbonne Universit\'e, 75005 Paris, France}
\affiliation{Laboratoire de Physique Th\'eorique de la Mati\`ere Condens\'ee, UMR7600 CNRS, Sorbonne Universit\'e, 75005 Paris, France}

\author{Thibault Bertrand}
\affiliation{Department of Mathematics, Imperial College London, South Kensington, London SW7 2AZ, United Kingdom}

\date{\today}

\begin{abstract}
\noindent 
\end{abstract}

\maketitle

\tableofcontents

\section{Lattice model and field theory derivation}\label{sec:Model}

\subsection{\textit{Partial exclusion process} augmented with detailed-balance breaking pairwise dynamics}\label{sec:LatticeModel}

We consider a lattice model of the \textit{partial exclusion process} \cite{Tailleur2008b, Floreani2021} on a periodic domain $\mathcal{L}$ with spacing $a$, augmented with the most minimal pairwise interaction that breaks detailed-balance while enforcing center-of-mass conservation (ensuring reciprocity) and bringing particles together (attractive). In deriving the continuum model, we consider only dimension $d=1$ dynamics, but extending the approach to arbitrary $d$ is straightforward. 

Let $n_i(t)$ denote the particle number at site $i$ and define the state $\bm{n}=\{n_i\}_{i\in \mathcal{L}}$. The model dynamics are then defined as follows:
\begin{itemize}
    \item[\circled{1}] Particles at site $i$ jump to site $i\pm 1$ with rate $\alpha_{i,\pm1}(\bm{n}) = D_r n_i (1- n_{i\pm1}/\kappa)$, where $D_r^{-1}$ sets the typical timescale associated with the diffusive motion and $\kappa$ is the carrying capacity at each site, assumed to be constant throughout the system;
    \item[\circled{2}] Two particles at sites $\{i-1, i+1\}$ simultaneously hop to site $i$ with rate $\beta_i(\bm{n}) = k_r n_{i-1} n_{i+1} \max(0,1 - n_i / (\kappa-1))$, where the partial exclusion due to the carrying capacity takes into account the $+2$ increase in particle number. 
\end{itemize} 
Note that as $n_i\leq \kappa$ the rates introduced above are always non-negative. Note the dynamical rule \circled{1} is exactly the partial exclusion process with carrying capacity $\kappa>1$. The addition of the second rule \circled{2} ensures broken detailed-balance and will drive phase separation: the coexistence of high- and low-density regions within the system.

\subsection{Mean field coarse-graining of lattice model}

The master equation for the state of this system can be written succinctly after defining $\bm{\Delta}_i=\delta_{i-1, j} - \delta_{i, j}$ where $\delta_{j, j'}$ is the Kronecker delta function:
\begin{align}
    \partial_t \mathbb{P}(\bm{n}(t)) =\sum_{i\in\mathcal{L}} &\bigg[-(\alpha_{i, -1} + \alpha_{i, +1} + \beta_{i-1} + \beta_{i+1})\mathbb{P}(\bm{n})\\
    & + \alpha_{i-1, +1}(\bm{n}+\bm{\Delta}_i)\mathbb{P}(\bm{n}+\bm{\Delta}_i)+ \alpha_{i+1, -1}(\bm{n}-\bm{\Delta}_{i+1})\mathbb{P}(\bm{n}-\bm{\Delta}_{i+1})\\
    & + \beta_i(\bm{n}+\bm{\Delta}_i-\bm{\Delta}_{i+1}) \mathbb{P}(\bm{n}+\bm{\Delta}_i-\bm{\Delta}_{i+1})\bigg]
\end{align}
where the coefficients are evaluated for the configuration $\bm{n}$ unless written otherwise. 

Having defined our lattice model, we now derive a Langevin equation for the particle density field through the construction of a Martin-Siggia-Rose dynamical action following the now standard approach \cite{Lefevre2007}. We first discretise time in to intervals $[t_j, t_{j+1}]$ of length $dt$. Our aim is to evaluate the dynamic distribution of particle distributions $\mathbb{P}(\bm{n}(t))$ in $[0, T]$. We do this by introducing the auxiliary jump process $J_{i}(t_j)= n_{i}(t_{j+1})-n_{i}(t_j)$ and the conjugated field $\tilde{n}_{i}(t_j).$ We express the dynamic distribution as a functional integral over the auxiliary field $\tilde{\bm{n}}(t)=\{\tilde{n}_{i}(t)\}$ in terms of an action $\mathcal{S}(\bm{n}(t),\tilde{\bm{n}}(t))$ as
\begin{equation}\label{eq:path-action}
    \mathbb{P}(\bm{n}(t)) = \frac{1}{Z} \int \mathcal{D}\tilde{\bm{n}} \: e^{-\mathcal{S}(\bm{n}(t), \tilde{\bm{n}}(t))}
\end{equation}
where $\mathcal{D}\tilde{\bm{n}} = \prod_{t_j}\prod_{i} d \tilde{n}_i$. We identify the action by considering the average of the arbitrary observable $\mathcal{O}(\tilde{\bm{n}}(t))$ over some period of time as a path integral over all possible trajectories. Following the Martin-Siggia-Rose approach, we write
\begin{equation}
    \langle \mathcal{O}\rangle = \frac{1}{Z}\bigg\langle \int \mathcal{D}\bm{n} \:\mathcal{O}(\bm{n}) \prod_{i}\delta \left(n_{i}(t_{j+1})-n_{i}(t_j) - J_{i}(t_j)\right)\bigg\rangle_{J_{i}}, 
\end{equation}
where $Z$ is chosen such that $\langle 1 \rangle=1.$ Here $\prod_i$ represents the product over all independent currents and times. By re-writing the Dirac-delta function as integrals over plane waves, we obtain the alternative formulation
\begin{equation}
    \langle \mathcal{O}\rangle = \frac{1}{Z} \int \mathcal{D}\bm{n} \int \mathcal{D}\tilde{\bm{n}}\:\mathcal{O}(\bm{n}) \big\langle \prod_{t_j}e^{\sum_{i}\tilde{n}_i(t_j)\left(n_{i}(t_{j+1})-n_{i}(t_j) - J_{i}(t_j)\right)}\big\rangle_{J_{i}}.
\end{equation}
The probability of a trajectory is thus 
\begin{equation}\label{eq:pprob}
    \mathbb{P}(\bm{n}(t)) = \frac{1}{Z}\int \mathcal{D}\tilde{\bm{n}}\:\big\langle \prod_{t_j} e^{-\sum_{i} \tilde{n}_{i}(t_j)(n_{i}(t_{j+1})-n_{i}(t_j) - J_{i}(t_j))}\big\rangle_{J_{i}}.
\end{equation}

Comparing Eq.\,\eqref{eq:pprob} to Eq.\,\eqref{eq:path-action}, we will determine the action $\mathcal{S}(\bm{n}(t), \tilde{\bm{n}}(t))$. Indeed, the generating function $\langle e^{\sum_i \tilde{n}_{i}J_{i}}\rangle_{J_i}$ (the only model specific part of the derivation) takes the form

\begin{eqnarray}
    \langle e^{\sum_i \tilde{n}_{i}J_{i}}\rangle_{J_i} & = & \left( 1 -dt \sum_{i}(\alpha_{i, -1} + \alpha_{i, +1} + \beta_{i-1} + \beta_{i+1}) \right)\\
    & & + dt \sum_{i}\left(\alpha_{i, +1} e^{\tilde{n}_{i+1}-\tilde{n}_{i}} +\alpha_{i, -1}e^{\tilde{n}_{i-1}-\tilde{n}_{i}} \right)\nonumber  +  dt\sum_{i} \beta_i e^{-\nabla^2 \tilde{n}_{i}} \\
    & & + \mathcal{O}(dt^2)\nonumber \\
    & \approx & e^{dt\,\mathcal{Q}(\bm{n},\tilde{\bm{n}}) }
\end{eqnarray}
where we have defined
\begin{equation}
        \mathcal{Q}(\bm{n},\tilde{\bm{n}})= \sum_{i\in\mathcal{L}}\sum_{\nu=\pm1}  \alpha_{i,\nu}(\bm{n}) \left(e^{\tilde{n}_{i+\nu}-\tilde{n}_{i}} -1\right) + \beta_{i}(\bm{n}) \left(e^{-\nabla^2 \tilde{n}_{i}} - 1\right).
\end{equation}

After transforming to continuous time and writing $\sum dt\rightarrow \int dt$, we arrive at the continuous-time dynamical action for our lattice model
\begin{equation}
    \mathcal{S} = \sum_{i\in\mathcal{L}} \int dt \left[\tilde{n}_{i}\partial_t n_{i} - \mathcal{Q}(\bm{n},\tilde{\bm{n}})\right].
\end{equation}
We now re-write the non-local terms (i.e. the $i\pm1$ terms appearing in the $i$-th term in the sum) in $\mathcal{Q}$ using Taylor expansions:
\begin{equation}
    n_{i+\nu} = n_{i} + a \nu\nabla_x n_i + \frac{a^2}{2} \nabla_x^2 n_i + \mathcal{O}(a^3) 
\end{equation}
for $\nu=\pm 1$ and similarly for $\tilde{n}_{i+\nu}$, where the (discrete) gradients are evaluated using centered finite differences. We also expand the exponential term in the action as
\begin{equation}
    e^{\tilde{n}_{i+\nu}-\tilde{n}_{i}} -1 = a\nu\nabla_x \tilde{n}_{i} + \frac{a^2}{2}\nabla^2_x \tilde{n}_{i} + \frac{a^2}{2} \left(\nabla_x \tilde{n}_{i}\right)^2 + \mathcal{O}(a^3)
\end{equation}
and similarly for $e^{-\nabla^2\tilde{n}_{i}} -1$.

After some algebra, we derive an expression for the action written entirely in local terms: 
\begin{eqnarray}
    \mathcal{S} = \sum_{i\in\mathcal{L}} \int dt \big[\tilde{n}_{i}\partial_t n_{i} - D_r a^2 n_i \left(1-\frac{n_i}{\kappa}\right)\nabla_x^2 \tilde{n}_i - D_r a^2 n_i \left(1-\frac{n_i}{\kappa}\right) \left(\nabla_x \tilde{n}_i\right)^2 + \frac{2D_r}{\kappa}a^2 n_i \nabla_x n_i \nabla_x \tilde{n}_i &\\
    + k_r a^2 \left(1-\frac{n_i}{\kappa-1}\right)n_i^2 \nabla_x^2 \tilde{n}_i - \frac{k_ra^4}{2} \left(1-\frac{n_i}{\kappa-1}\right)n_i^2 (\nabla^2_x\tilde{n}_i)^2+ \dots& \big]\nonumber,
\end{eqnarray}
where we have kept the leading-order terms of orders $\mathcal{O}(\tilde{n}_i)$ and $\mathcal{O}(\tilde{n}^2_i)$ for both the diffusion and interaction terms. These will represent the leading-order (in spatial gradients) deterministic and stochastic contributions to the effective density equation, respectively. Note that, while not written here (for brevity), higher-order terms (in spatial gradients) contribute to the $\nabla^4$ term discussed below. 

To derive a continuum action, one may look to take $a\rightarrow 0$. However, in our model, this sets the range of interactions, thus the particle number itself would need to diverge to produce a non-trivial continuum limit. Following similar coarse-graining approaches for lattice models with simultaneous hopping interactions \cite{Han2024}, we fix the lattice spacing $a=1$ and neglect higher-order gradient terms on the assumption that only the leading-order terms in spatial gradients contribute to the dynamics over this lengthscale. This is motivated later through comparison of numerical simulations of the lattice (particle based) model to the effective continuum model. Doing so, we arrive at the continuum action for the field $n(x, t)$ (and its conjugate $\tilde{n}(x, t)$) of the form 
\begin{align}
    \mathcal{S} = \int dx \int dt \bigg[&\tilde{n}\partial_t n - D_r n \left(1-\frac{n}{\kappa}\right)\nabla_x^2 \tilde{n} - D_r n \left(1-\frac{n}{\kappa}\right) \left(\nabla_x \tilde{n}\right)^2 + \frac{2D_r}{\kappa} n \nabla_x n \nabla_x \tilde{n}     \\
    &+ k_r\left(1-\frac{n}{\kappa-1}\right)n^2 \nabla_x^2 \tilde{n} - \frac{k_r}{2} \left(1-\frac{n}{\kappa-1}\right)n^2 (\nabla^2_x\tilde{n})^2 \bigg]\nonumber,
\end{align}
from which we read off the Langevin equation for $n(x, t)$ that crucially contains two sources of noise: 
\begin{equation}
    \partial_t n(x, t) = \partial_x\left[D_r- k_r\left(2n - \frac{3 n^2}{\kappa-1} \right)\right]\partial_x n + \partial_x \left[\sqrt{2 D_r n \left(1-\frac{n}{\kappa}\right)} \Lambda(x, t)\right] + \partial_x^2 \left[\sqrt{k_r n^2 \left(1-\frac{n}{\kappa-1}\right)}\eta(x, t)\right].
\end{equation}
We then linearise around the average $\bar{n}$, defining the density fluctuations as $\phi = n(x, t)-\bar{n}$ and deriving

\begin{align}
    \partial_t \phi(x, t) = D_0 \partial_x^2 \phi + \sqrt{2D_1} \partial_x\Lambda(x, t) + \sqrt{A}\partial^2_x\eta(x, t),
\end{align}
where the stochastic terms $\Lambda(x, t)$ and $\xi(x, t)$ are zero-mean, unit variance Gaussian white noise processes and we have defined \begin{subequations}
\begin{align}
D_0 &= D_r - k_r \left( 2 \bar{n} - \frac{3\bar{n}^2}{\kappa-1} \right),\quad D_1 = D_r\bar{n}\left(1 - \frac{\bar{n}}{\kappa}\right)~~\mathrm{and} ~~A = k_r\bar{n}^2\left(1 - \frac{\bar{n}}{\kappa-1}\right). 
\end{align}
\end{subequations}
We remark here that the decision to neglect higher-order terms restricts the validity of the dynamical description to long-wavelength dynamics: when calculating observables below, we impose a minimum wavelength (set by the lattice spacing to unity) to avoid divergences at higher Fourier modes $q$.

One can extend the methodology employed to derive the coefficient for the surface tension term, that is, the fourth-order gradient term linear in $\phi$. We omit this lengthy algebra from the derivation above to improve readability, but the framework is the same. The result is a dynamics 
\begin{equation}
    \partial_t \phi(x, t) = D_0 \partial_x^2 \phi - \gamma\partial_x^4\phi + \sqrt{2D_1} \partial_x\Lambda(x, t) + \sqrt{A}\partial^2_x\eta(x, t), \label{eqSI:phidot}
\end{equation}
where $\gamma = \frac{1}{12}\left[D_r  + k_r \left( 14 \bar{n} - \frac{15\bar{n}^2}{\kappa-1} \right) \right].$ As noted in the main text, the dynamics in Eq.\,\eqref{eqSI:phidot} cannot be understood as those of Model B in the Gaussian approximation \cite{Hohenberg1977} with a linearly re-scaled temperature due to the presence of the $\sqrt{A}$ term. Indeed, in Fourier space, it is Model B dynamics with a $q$-dependent effective temperature. This signifies that the equilibrium fluctuation-dissipation relation for Model B does not hold for these dynamics.  

\subsection{Numerical verification at high density}

\begin{figure}
    \centering
    \includegraphics[width=0.8\linewidth]{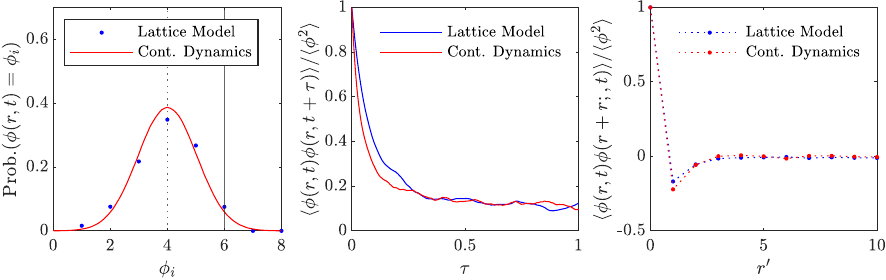}
    \caption{\textit{Comparison of simulations to coarse-graining ---} We set $\bar{n}=4$ and $n_j=6$ and compare the spatial and temporal correlation functions in simulations of the two dynamics (discrete lattice model and continuous field dynamics). We find quantitative agreement between the distribution of particle numbers, the relaxation time of the two-time correlation function and the relaxation length for the two-space correlation function (which is around 1 lattice space). }
    \label{fig:latt_nums1}
\end{figure}

\begin{figure}
    \centering
    \includegraphics[width=0.8\linewidth]{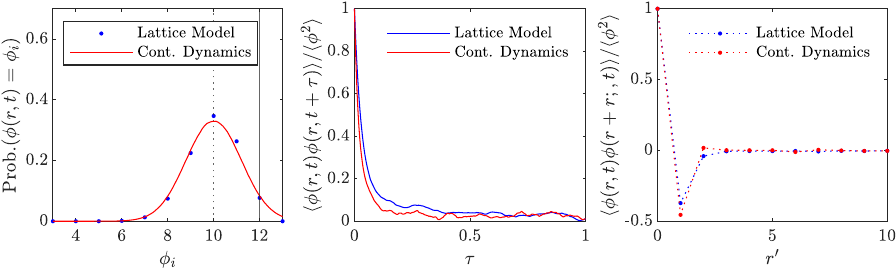}
    \caption{\textit{Comparison of simulations to coarse-graining ---} We set $\bar{n}=10$ and $n_j=12$ and again compare the spatial and temporal correlation functions in simulations of the two dynamics (discrete lattice model and continuous field dynamics). We find stronger agreement for the increased particle number and carrying capacity: we argue that this is because the analytics can only be shown to capture the mean field behaviour, which is best approximated in the limit of large particle number and carrying capacity per site.  
    }
    \label{fig:latt_nums2}
\end{figure}

To justify the value in our mean-field coarse-graining procedure, we compare numerical simulations of the two dynamics. The results are given in Figures \ref{fig:latt_nums1} and \ref{fig:latt_nums2}. We set the rates to unity, $D_r=k_r=1$ and vary $\bar{n}$ and $n_j$, which play the same role in the two descriptions, namely the mean and maximum density. In each case, we see quantitative agreement in the distribution in values for the particle number field (left panel in each figure). We further find good agreement in the spatial and temporal correlation functions for the field (plotted in the remaining panels) when the mean density is close to the jamming density: this is precisely the regime that we are interested in and so we conclude that the derived continuous descriptions is suitable for analysing the transport properties of our system. 

\section{Calculating transport properties for probe particle in fluctuating field}

We now consider the dynamics of a probe particle diffusing in the field dynamics detailed above for the continuum model. The particle-field coupling is reciprocal with coupling strength $h$, for which it is well understood that equilibrium dynamics for the field (i.e. dynamics satisfying the fluctuation-dissipation relation) give a negative leading order correction to the bare diffusivity of the probe \cite{Dean2011, Demery2011}. Here, we demonstrate that sufficiently strong non-equilibrium fluctuations (stemming from detailed-balance breaking microscopic interactions) can drive an enhancement in the self-diffusion coefficient of the probe particle (to leading order in the coupling strength). The derivation follows closely the one described in \cite{Dean2011, Demery2011, Demery_2014}, for which results have been obtained for field dynamics with fluctuations at a different temperature, trivially breaking the fluctuation-dissipation relation. Here we consider the effect of gradient terms breaking this relation, for which our continuum model detailed above is a example case. 

In what follows, the composition of two operators $A(x, x')$ and $B(x, x')$ is written as $AB(x, x')=  \int dx'' A(x, x'')B(x'', x')$ and similarly for three or more operators in succession. We consider the following coupled dynamics for a probe particle at position $\bm{y}(t)$ and a field $\phi(\bm{x}, t)$: the particle evolves under the dynamics
\begin{equation}
    \dot{\bm{y}}(t) = \kappa_y\bm{f}+h \kappa_y\nabla K \phi[\bm{y}(t)] + \sqrt{\kappa_y}\bm{\eta}_{y}(t),
\end{equation}
where $h$ denotes the coupling constant between particle and field (which we will assume to be small in order to treat the coupling perturbatively), $\kappa_y$ represents the bare motility of the probe particle in the background fluid, $\nabla$ is a standard gradient operator and $K$ determines the nature of the coupling with the field. 

We consider the generalised dynamics for the field of the form 
\begin{equation}
    \partial_t \phi(\bm{x}, t) = - \kappa_\phi R \Delta \phi + h \kappa_\phi R K[\bm{x}-\bm{y}(t)] + \sqrt{\kappa_\phi}\xi(\bm{x}, t),
\end{equation}
where the operator $\Delta$ describes its diffusive behaviour and the zero-mean noise $\xi(\bm{x}, t)$ satisfies the correlation function
\begin{equation}
    \langle \xi(\bm{x}, t) \xi(\bm{x}', t')\rangle = 2T_\phi R(\bm{x}-\bm{x}')\Upsilon(\bm{x}-\bm{x}')\delta(t-t').
\end{equation}
This is suitable to capture the field dynamics studied above, but we leave it general here for now. Here $R$ determines whether the field dynamics are conserved (Model B) or not (Model A). As an example, a point particle and a field with model A dynamics would lead us to choose $R=K=\delta(\bm{x}-\bm{x}')$ and $\Delta = \left(m^2 - \nabla_{\bm{x}}^2\right)\delta(\bm{x}-\bm{x}')$.

\subsection{Effective Diffusion Equation}

We integrate the dynamical equation for the field, assuming $\phi=0$ at time $t=0$, giving
\begin{equation}
    \phi(\bm{x}, t) = \int_{-\infty}^t e^{-\kappa_\phi(t-s)R \Delta} \left[h\kappa_\phi RK\left[\bm{x}-\bm{y}(s)\right]+\sqrt{\kappa_\phi}\xi(\bm{x}, s)\right]ds.
\end{equation}
The infinite lower integration bound signifies that the system is in a stationary state, having forgotten about its initial configuration. From this solution, we can derive an effective diffusion equation for the particle, namely
\begin{equation}\label{eq:eom-y}
    \dot{\bm{y}}(t) = \kappa_y\bm{f} + \sqrt{2\kappa_y}\bm{\eta}_{y}(t) + h \kappa_y\nabla K \int_{-\infty}^t e^{-\kappa_\phi(t-s)R \Delta}\left[h \kappa_\phi R K[\bm{y}(t)-\bm{y}(s)]+ \sqrt{\kappa_\phi}\xi(\bm{y}(t), s)\right]ds.
\end{equation}
This dynamical equation for $\bm{y}(t)$ can be written in the form 
\begin{equation}
    \dot{\bm{y}}(t) = \kappa_y\bm{f}+ \sqrt{2\kappa_y}\bm{\eta}_{y}(t) + \int_{-\infty}^t \bm{F}(\bm{y}(t)-\bm{y}(s), t-s) ds + \Xi(\bm{y}(t), t), 
\end{equation}
where $\Xi(\bm{x}, t)$ is a Gaussian noise with correlation function 
\begin{equation}
    \left\langle \Xi(\bm{x}, t)\Xi(\bm{x}', t')^T\right\rangle = \bm{G}(\bm{x}-\bm{x}', t-t').
\end{equation}
The two relevant functions are defined as 
\begin{align}
    \bm{F}(\bm{x}, t) &= h^2\kappa_y\kappa_\phi \nabla K e^{-\kappa_\phi t R \Delta} R K(\bm{x})\\
    \bm{G}(\bm{x}, t) &= -T_\phi h^2 \kappa_y^2\nabla \nabla^T K^2e^{-\kappa_\phi|t|R \Delta}\Upsilon\Delta^{-1}(\bm{x}).
\end{align}
with Fourier transforms
\begin{align}
    \tilde{\bm{F}}(\bm{q}, t) &= i h^2\kappa_y\kappa_\phi \bm{q} \tilde{K}(\bm{q})^2 \tilde{R}(\bm{q})e^{-\kappa_\phi\tilde{R}(\bm{q})\tilde{\Delta}(\bm{q})t}\\
    \tilde{\bm{G}}(\bm{q}, t) &= T_\phi h^2\kappa_y^2 \bm{q} \bm{q}^T\tilde{K}(\bm{q})^2 \tilde{\Upsilon}(\bm{q}) \tilde{\Delta}^{-1}(\bm{q})e^{-\kappa_\phi\tilde{R}(\bm{q})\tilde{\Delta}(\bm{q})|t|}.
\end{align}

\subsection{Field Theory for EOM}
The dynamics of Eq.\,\eqref{eq:eom-y} can be mapped to a field theory with action $S[\bm{y}, \conj] = S_0[\bm{y}, \conj]+S_{\rm int}[\bm{y}, \conj]$, where we have defined the bare action 
\begin{equation}
    S_0[\bm{y}, \tilde{\bm{y}}] = -i \int \conj(t)\cdot\left[\dot{\bm{y}}(t) - \kappa_y\bm{f}\right] dt + D_y \int |\tilde{\bm{y}}(t)|^2dt
\end{equation}
where $D_y = \kappa_y  T$ and the remaining (interaction) terms in the action
\begin{equation}
    S_{\rm int}[\bm{y}, \conj] = i \int \conj \cdot \bm{F}(\bm{y}(t)-\bm{y}(t'), t-t')\theta(t-t') dtdt' + \int \conj(t)\cdot \bm{G}(\bm{y}(t)-\bm{y}(t'), t-t')\conj(t')\theta(t-t')dt dt'
\end{equation}
where we have introduced the conjugate field $\conj(t)$ and $\theta(t)$ is the Heaviside function. We work in the It\^o convention.

\subsection{Computing Averages with the Action $S_0$}

We compute averages of observables, such as $\langle \mathcal{O}(\bm{y})\rangle$ with respect to the full action by calculating
\begin{equation}
    \langle \mathcal{O}(\bm{y})\rangle = \int \mathcal{O}e^{-S[\bm{y}, \conj]}d \bm{y} d \conj/ Z,
\end{equation}
where $Z$ is such that $\langle 1\rangle = 1.$
We denote the average with respect to the bare action with the notation $\langle \mathcal{O}\rangle_0$ and derive the relation between the two averages:
\begin{equation}\label{eq:mean_rel}
    \langle \mathcal{O}[\bm{y}]\rangle = \frac{\langle \mathcal{O}[\bm{y}] \exp(-S_{\rm int}[\bm{y}, \conj])\rangle_0}{\langle \exp(-S_{\rm int}[\bm{y}, \conj])\rangle_0}\simeq \frac{\langle \mathcal{O}[\bm{y}] (1-S_{\rm int}[\bm{y}, \conj])\rangle_0}{\langle 1-S_{\rm int}[\bm{y}, \conj]\rangle_0}.
\end{equation}
By Wick's theorem, we require only the two first moments to compute all averages. We compute these as 
\begin{align}
    \langle \bm{y}(t)\rangle_0 &= \kappa_y\bm{f} t\\
    \langle \conj(t)\rangle_0 &= \bm{0}\label{eq:meanconj}\\
    \langle \conj(t)\conj(t')^T\rangle_0 &= \bm{0}\label{eq:varconj}\\
    \langle \bm{y}(t)\conj(t')^T\rangle_0 &= i\chi_{[0, t)}(t')\\
    \langle \left[\bm{y}(t)-\kappa_y\bm{f}t\right]\left[\bm{y}(t')-\kappa_y\bm{f}t'\right]^T\rangle_0&=2\kappa_yT \max(t, t')
\end{align}
where $\chi_A(t)$ is the characteristic function for the interval $A$: $1$ is $t\in A$ and zero otherwise. 

The following result, which can be derived from Wick's theorem, details how to re-write these terms using correlators that no longer contain an exponential:
\begin{equation}\label{eq:wicks}
    \left\langle \prod_{j=1}^n \mathcal{O}_j e^{i \bm{q}\cdot \bm{y}}\right\rangle_0 = e^{i \bm{q}\cdot \langle \bm{y}\rangle_0-\frac{1}{2} \bm{q}^T\langle \bm{y}\bm{y}^T\rangle_0\bm{q} } \sum_{J\subset N} i^{|J|} \left(\prod_{j\in J} \bm{q} \cdot\langle \mathcal{O}_j \bm{y}\rangle_0 \left\langle \prod_{k\notin J}\mathcal{O}_k\right\rangle_0\right)
\end{equation}
where $N$ is the set $\{1, \dots, n\}$ and the sum over $J$ denotes the sum over all subsets of $N$. 

We first consider $\langle S_{\rm int}[\bm{y}, \conj]\rangle_0$: the calculation simplifies after re-writing $\bm{F}(x, t)=\frac{1}{(2\pi)^d}\int d\bm{q}\tilde{\bm{F}}e^{i\bm{q}\cdot\bm{x}}$, thus shifting all the spatial dependence in to the factor $e^{i\bm{q}\cdot \bm{x}}.$  By Eqs.\,\eqref{eq:meanconj}, \eqref{eq:varconj} and \eqref{eq:wicks}, we evaluate
\begin{equation}
    \langle \conj(t)e^{i\bm{q}\cdot\left[\bm{y}(t)-\bm{y}(t')\right]}\rangle_0=
    \langle \conj(t)\conj(t')^Te^{i\bm{q}\cdot\left[\bm{y}(t)-\bm{y}(t')\right]}\rangle_0=\bm{0}
\end{equation}
and thus $\langle S_{\rm int}\left[\bm{y}, \conj\right]\rangle_0=0.$ The remaining averages require a definition of the observable, which we detail below. 

\subsection{Computing the Mobility}

We define the effective mobility of a tracer particle by applying a constant external force $\bm{f}$ and evaluating 
\begin{equation}\label{eq:mob_def}
    \kappa_{\rm eff}(\bm{f}) = \lim_{t\rightarrow\infty}\frac{\langle \bm{y}(t)\rangle}{\bm{f}t}.
\end{equation}
The required observable is thus $\mathcal{O}=\bm{y}(t)$, resulting in the term $\langle \bm{y}(t)S_{\rm int}[\bm{y}, \conj]\rangle_0$. As above, we compute the two required terms (for $t'>t''$):
\begin{align}
    \langle \bm{y}(t)\conj(t')^T e^{i \bm{q}\cdot[\bm{y}(t')-\bm{y}(t'')]}\rangle_0 &= i \chi_{[0, t)}(t')e^{i\kappa_y\bm{q}\cdot \bm{f}(t'-t'')- \kappa_yT \bm{q}^2 |t'-t''|}\\
    \langle \bm{y}(t)\conj(t')^T\conj(t'')e^{i \bm{q}\cdot[\bm{y}(t')-\bm{y}(t'')]}\rangle_0 &= -i \chi_{[0, t)}(t')\bm{q}e^{i\kappa_y\bm{q}\cdot \bm{f}(t'-t'')- \kappa_yT \bm{q}^2 |t'-t''|}.
\end{align}
The resulting expression for the required average (for long times) after integrating over the interaction times $t'$ and $t''$ gives
\begin{equation}
    \langle \bm{y}(t) S_{\rm int}[\bm{y}, \conj]\rangle_0 \sim \frac{t \kappa_y^2 \bm{f} h^2}{\kappa_\phi} \int \frac{d\bm{q}}{(2\pi)^d} \frac{q_{\parallel}^2 |\tilde{K}(\bm{q})|^2 \left[\kappa_\phi\tilde{R}(\bm{q})\tilde{\Delta}(\bm{q})+\kappa_yT\bm{q}^2\right]}{\tilde{\Delta}(\bm{q})\left(\left[\kappa_\phi\tilde{R}(\bm{q})\tilde{\Delta}(\bm{q})+\kappa_yT\bm{q}^2\right]^2 + \kappa_y^2\left[\bm{f}\cdot \bm{q}\right]^2\right)}
\end{equation}

By Eq.\,\eqref{eq:mean_rel} and our expression for the mobility in Eq.\,\eqref{eq:mob_def}, we derive
\begin{equation}
    \kappa_{\rm eff}(\bm{f}) = \kappa_y \left[1 - \frac{h^2 \kappa_y}{\kappa_\phi}\int \frac{d\bm{q}}{(2\pi)^d} \frac{q_{\parallel}^2 |\tilde{K}(\bm{q})|^2 \left[\kappa_\phi\tilde{R}(\bm{q})\tilde{\Delta}(\bm{q})+\kappa_yT\bm{q}^2\right]}{\tilde{\Delta}(\bm{q})\left(\left[\kappa_\phi\tilde{R}(\bm{q})\tilde{\Delta}(\bm{q})+\kappa_yT\bm{q}^2\right]^2 + \kappa_y^2\left[\bm{f}\cdot \bm{q}\right]^2\right)}\right].
\end{equation}
In the limit of vanishing forcing, we evaluate the bare friction coefficient (that is, the one that appears in a classical Einstein relation) as
\begin{equation}\label{eq:keff_res}
    \kappa_{\rm eff}(\bm{f}\ll 1) = \kappa_y \left[1 - \frac{h^2\kappa_y}{\kappa_\phi}\int \frac{d\bm{q}}{(2\pi)^d} \frac{q_{\parallel}^2 |\tilde{K}(\bm{q})|^2 }{\tilde{\Delta}(\bm{q})\left[\kappa_\phi\tilde{R}(\bm{q})\tilde{\Delta}(\bm{q})+\kappa_yT\bm{q}^2\right]}\right].
\end{equation}
We remark here that no dependence on $\tilde{\Upsilon}(\bm{q})$ appears, indicating that the result for the effective friction coefficient does not change when we add our source of fluctuations (those breaking the fluctuation dissipation relation) in the field dynamics. Indeed, this result has been derived previously the case of Model B. 

\subsection{Computing the Self-Diffusion}

Next, we define the effective self-diffusion coefficient for the probe particle as
\begin{equation}
    D_{\rm eff} = \lim_{t\rightarrow \infty}\frac{\langle \left[\bm{y}(t)-\langle \bm{y}(t)\rangle\right]^2\rangle}{2dt}.
\end{equation}
Using $\langle \bm{y}(t)\rangle=\kappa_y\bm{f}t$, we see that we need to calculate $\langle \left[\bm{y}(t)-\kappa_y\bm{f}t\right]^2S_{\rm int}(\bm{y}, \conj)\rangle_0$. We thus calculate (again for $t'>t''$)
\begin{align}
    \langle [\bm{y}(t)-\kappa_y\bm{f}t]^2\conj(t')^Te^{i\bm{q}\cdot (\bm{y}(t')-\bm{y}(t''))}\rangle_0 &= -4\kappa_yT\bm{q}^T\chi_{(0, t)}(t')(t'-t'')e^{i\kappa_y\bm{q}\cdot\bm{f}(t'-t'')-\kappa_yT\bm{q}^2|t'-t''|}\\
    \langle [\bm{y}(t)-\kappa_y\bm{f}t]^2\conj(t')^T\bm{p}(t'')e^{i\bm{q}\cdot (\bm{y}(t')-\bm{y}(t''))}\rangle_0 &= -2\chi_{(0, t)} (t')\left[2\kappa_yT\bm{q}^2(t'-t'')-\chi_{(0, t)}(t'')\right]e^{i\kappa_y\bm{q}\cdot \bm{f}(t'-t'')-\kappa_yT\bm{q}^2|t'-t''|}
\end{align}
from which we derive (in the long time limit)
\begin{align}
    \langle [\bm{y}-\kappa_y\bm{f}t]^2 &S_{\rm int}[\bm{y}, \conj ]\rangle_0 \simeq \nonumber\\
    &    {2t h^2 \kappa_y^2} \int \frac{d\bm{q}}{(2\pi)^d}\frac{|\bm{q}|^2|\tilde{K}(\bm{q})|^2\left[T_\phi T |\bm{q}|^2\tilde{\Upsilon}(\bm{q}) + (2T-T_\phi \tilde{\Upsilon}(\bm{q}))\kappa_\phi\tilde{R}(\bm{q})\tilde{\Delta}(\bm{q})\right] \left[\left[\kappa_\phi\tilde{R}(\bm{q})\tilde{\Delta}(\bm{q})+\kappa_y T\bm{q}^2\right]^2 - 3 \kappa_y^2 (\bm{f}\cdot\bm{q})^2\right]}{\tilde{\Delta}(\bm{q})\left[\left[\kappa_\phi\tilde{R}(\bm{q})\tilde{\Delta}(\bm{q})+\kappa_y T\bm{q}^2\right]^2 + \kappa_y^2(\bm{f}\cdot \bm{q})^2\right]^2}.
\end{align}
Finally, we arrive at the expression for the effective diffusion coefficient for the probe particle subject to the driving force $\bm{f}$:
\begin{equation}
    D_{\rm eff}(\bm{f}) = D_y - \frac{h^2\kappa^2_y }{d}\int \frac{d\bm{q}}{(2\pi)^d}\frac{|\bm{q}|^2|\tilde{K}(\bm{q})|^2\left[T_\phi T |\bm{q}|^2\tilde{\Upsilon}(\bm{q}) + (2T-T_\phi \tilde{\Upsilon}(\bm{q}))\kappa_\phi\tilde{R}(\bm{q})\tilde{\Delta}(\bm{q})\right] \left[\left[\kappa_\phi\tilde{R}(\bm{q})\tilde{\Delta}(\bm{q})+\kappa_y T\bm{q}^2\right]^2 - 3 \kappa_y^2 (\bm{f}\cdot\bm{q})^2\right]}{\tilde{\Delta}(\bm{q})\left[\left[\kappa_\phi\tilde{R}(\bm{q})\tilde{\Delta}(\bm{q})+\kappa_y T\bm{q}^2\right]^2 + \kappa_y^2(\bm{f}\cdot \bm{q})^2\right]^2}.
\end{equation}
In the limit of zero-driving force, we recover the bare diffusion coefficient due to interactions with the field:
\begin{equation}\label{eq:Deff_res}
    D_{\rm eff}(\bm{f}\rightarrow 0) = D_y - \frac{h^2\kappa^2_y }{d}\int \frac{d\bm{q}}{(2\pi)^d}\frac{|\bm{q}|^2|\tilde{K}(\bm{q})|^2\left[\kappa_yT_\phi T |\bm{q}|^2\tilde{\Upsilon}(\bm{q}) + (2T-T_\phi \tilde{\Upsilon}(\bm{q}))\kappa_\phi\tilde{R}(\bm{q})\tilde{\Delta}(\bm{q})\right] }{\tilde{\Delta}(\bm{q})\left[\kappa_\phi\tilde{R}(\bm{q})\tilde{\Delta}(\bm{q})+\kappa_y T|\bm{q}|^2\right]^2}.
\end{equation}
Unlike the effective motility evaluated above, the effective self-diffusion does depend on $\tilde \Upsilon(\bm{q})$, implying that the breaking of FDT for the field dynamics does impact the diffusion of the probe particle. Interestingly, this result exactly matches that of Eq.\,(43) in Ref.\,\cite{Demery2011} after replacing $T_{\phi}$ in their work with $T_\phi \Upsilon(\bm{q})$. While in hindsight, one could have expected this result from those Ref.\,\cite{Demery2011}, we provide the full calculation here for clarity. It is instructive to re-write Eq.\,\eqref{eq:Deff_res} after substituting for $T$ through $D_y=\kappa_y T$:
\begin{equation}\label{eq:Deff_res}
    D_{\rm eff}(\bm{f}\rightarrow 0) = D_y - \frac{h^2\kappa_y }{d}\int \frac{d\bm{q}}{(2\pi)^d}\frac{|\bm{q}|^2|\tilde{K}(\bm{q})|^2\left[\kappa_yT_\phi D_y |\bm{q}|^2\tilde{\Upsilon}(\bm{q}) + (2D_y-\kappa_yT_\phi \tilde{\Upsilon}(\bm{q}))\kappa_\phi\tilde{R}(\bm{q})\tilde{\Delta}(\bm{q})\right] }{\tilde{\Delta}(\bm{q})\left[\kappa_\phi\tilde{R}(\bm{q})\tilde{\Delta}(\bm{q})+D_y|\bm{q}|^2\right]^2}.
\end{equation}

\subsection{Results for Model of Section \ref{sec:Model}}

Recalling the dynamics of Eq.\,\eqref{eqSI:phidot}, we identify the functional forms relating to the model studied in the previous section: for $\kappa_y=\kappa_\phi=1$ we identify $\tilde{K}(q)=1$ and $\tilde{R}(q) = q^2$. Then defining $\mu_\phi T_{\rm eff}(q) = T_\phi\tilde{\Upsilon}(q) = D_1 + \frac{A}{2}q^2$ and noting $\tilde{\Delta}(q) = D_0 + \gamma q^2$, the final results for the mobility and self-diffusion take the simplified forms
\begin{equation}
    \kappa_{\rm eff}(f \ll 1) = \left[1 - h^2 \int_{|\bm{q}|<\frac{2\pi}{a}} \frac{d^d\bm{q}}{(2\pi)^d} \frac{1}{\tilde{\Delta}(\bm{q})\left[ D_y+\tilde{\Delta}(\bm{q})\right]}\right]
\end{equation}
and
\begin{equation}\label{eq:selfDiff}
    {D_{\rm eff}}= D_y 
    - \frac{h^2}{d} \int \frac{d^d\mathbf{q}}{(2\pi)^d} \frac{D_y\mu_\phi T_{\rm eff}(\mathbf{q}) + (2D_y-\mu_\phi T_{\rm eff}(\mathbf{q}))\tilde{\Delta}(\mathbf{q})}{\tilde{\Delta}(\mathbf{q})(D_y + \tilde{\Delta}(\mathbf{q}))^2}.
\end{equation}
We compare the result for the self-diffusion to numerical simulations in Figure 3 of the main text, finding good agreement for particle-field coupling of values up to $h=1$. Note that the expression for $D_{\rm eff}$ quoted in the main text omits the tilde from $\tilde{\Delta}$ in Eq.\eqref{eq:selfDiff}, but the result is the same.

\section{Details of coupled particle-field simulations}

Following the numerical approach of Ref.\,\cite{Demery2011}, we solve the dynamics for the field in Fourier space but keep those of the particle in real space, working in dimension $d=1$. As above, we define our Fourier transform as
\begin{equation}
    \tilde{\phi}(q, t) = \int_{-\infty}^\infty dx\: e^{-iqx} \phi(x, t),\quad \phi(x, t) = \frac{1}{2\pi}\int_{-\infty}^\infty dq \: e^{iqx}\tilde{\phi}(q, t).
\end{equation}

The field's dynamics in Fourier space take the form
\begin{equation}\label{eq:fielddyn}
    \partial_t \tilde\phi(q, t) = - D_0 q^2 \tilde{\phi} - \gamma q^4  \tilde{\phi} + h \mu_\phi q^2 e^{-i q y(t)}  + \sqrt{\mu_\phi} {\xi}(q, t)
\end{equation}
where we have the following correlator for the transformed noise
\begin{equation}
    \langle \xi(q, t) \xi(q', t')\rangle  = \left[2 T q^2+ A'q^4 \right]\delta(q+q')\delta(t-t').
\end{equation}

The dynamics for the probe particle then take the form 
\begin{align}\label{eq:probedyn}
    \dot{y}(t) &= h\kappa_y \partial_y \phi(y(t), t) + \sqrt{\kappa_y}\eta(t)
\end{align}
where the zero-mean white noise $\eta(t)$ has the correlator
\begin{equation}
    \langle \eta(t)\eta(t')\rangle = 2T\delta (t-t').
\end{equation}

We solve the coupled equations using the second-order method of Ref.\,\cite{Branka1999}. We discretise the wave-numbers in Fourier space as $q_i=\{\frac{2\pi}{N_q}, \frac{4\pi}{N_q}, \dots, 2\pi\}.$ In simulations, we choose $\kappa_y=1$ and $N_q=10$.


%